\documentclass[12pt]{article}

\def\D{\Delta}

\def\L{\Lambda}
\def\l{\lambda}
\def\S{\Sigma}
\def\G{\Gamma}
\def\g{\gamma}
\def\e{\epsilon}
\def\ve{\varepsilon}
\def\s{\sigma}

\def\a{\alpha}

\def\m{\mu}
\def\n{\nu}

\def\s{\sigma}

\def\e{\epsilon}
\def\ve{\varepsilon}
\def\dim{\textrm{dim}}
\def\det{\textrm{det}}

\def\E{\textrm{erf}}
\newcommand{\be}{\begin{equation}}
\newcommand{\ee}{\end{equation}}
\newcommand{\bea}{\begin{eqnarray}}
\newcommand{\eea}{\end{eqnarray}}

\begin{document}

\begin{center}
\bf{\large  EFFECTIVE ACTIONS FOR REGGE STATE-SUM MODELS OF QUANTUM GRAVITY}
\end{center}

\bigskip
\begin{center}
{\large Aleksandar Mikovi\'c} 

\bigskip
{\it Departamento de Matem\'atica  \\
Universidade Lus\'ofona de Humanidades e Tecnologias\\
Av. do Campo Grande, 376, 1749-024 Lisboa, Portugal}\\
and\\
{\it Grupo de Fisica Matem\'atica da Universidade de Lisboa\\
Av. Prof. Gama Pinto, 2, 1649-003 Lisboa, Portugal}\\
\end{center}

\centerline{E-mail: amikovic@ulusofona.pt}

\bigskip
\bigskip
\begin{quotation}
\noindent\small{We study the semiclassical expansion of the effective action for a Regge state-sum model and its dependence on the choice of the path-integral measure and the spectrum of the edge lengths. If the positivity of the edge lengths is imposed in the effective action equation, we find that the semiclassical expansion is not possible for the power-law measures, while the exponential measures allow the semiclassical expansion. Furthermore, the exponential measures can generate the cosmological constant term in the effective action as a quantum correction, with a naturally small value of the cosmological constant. We also find that in the case of a discrete-length spectrum, the semiclassical expansion is allowed only if the spectrum gap is much smaller than the Planck length. }\end{quotation}

\bigskip
\bigskip
\noindent{\bf{1. Introduction}}

\bigskip
\noindent A natural way to define the path integral for General Relativity (GR) is to discretize GR by using a spacetime triangulation and to use Regge calculus, for a review and references see \cite{h}. The main goal of quantum Regge calculus (QRC) is to define the GR path integral
\be Z_0 = \int {\cal D}g \exp \left({i\over l_P^2 }\int_M \sqrt{|det\,g|} \, R(g) \, d^4 x\right)\,,\ee
where $g$ is a metric on a 4-manifold $M$, $R(g)$ is the scalar curvature and $l_P$ is the Planck length, as an appropriate limit of a discrete path integral
\be Z_R  = \int_{D_E} \prod_{\e=1}^E dL_\e  \,\m (L) \exp\left({i\over l_P^2 } \sum_{\D =1}^F A_\D (L)\, \delta_\D (L) \right)\,, \label{rpi}\ee
where $\e$ denotes an edge of a triangulation of $M$, $L_\e$ denotes an edge length and
$$ S_R (L) = \sum_{\D =1}^F A_\D (L)\, \delta_\D (L)\,, $$
is the Regge action. Here $F$ is the number of triangles in the triangulation, $A_\D$ is the area of a traingle $\D$ and $\delta_\D$ is the corresponding deficit angle. The integration region $D_E$ is a subset of ${\bf R}_+^E$ where the triangle inequalities for the edge lengths hold and $E$ is the number of edges. 

The measure $\m(L)$ is chosen in quantum Regge calculus as
\be \m (L) = \prod_{\e=1}^E (L_\e)^p \,\prod_{\s=1}^N (V_\s (L))^q \, e^{-\l V_4(L)}\,,\label{qrcm}\ee
where $\s$ denotes  a 4-simplex of the triangulation, $V_\s$ is the 4-volume of $\s$ and $V_4 (L)$ is the triangulation 4-volume \cite{h}. The parameters $p,\l$ and $q$ are real numbers, and the convergence of $Z_R$ requires that $p,q \ge 0$ and $\l > 0$.

In order to define a smooth-manifold limit of $Z_R$, one can adopt the Wilsonian approach, see \cite{h,cdt}. In this approach $\kappa = i/l_P^2$ is allowed to vary and it will play the same role as the temperature in critical phenomena. One then looks for the critical points of $Z_R (\kappa)$, and if the right kind of critical points exists, then in the vicinity of such a critical point the correlation length diverges, and the discrete theory would be described by a quantum field theory which should correspond to quantum GR on a smooth manifold. Note that in the Wilsonian approach the number of the edges $E$ can remain finite, although it has to be a large number. Also $\kappa$ is taken to be variable because it is assumed that the Newton constant $G_N$ is  a running coupling constant, i.e. depends on a characteristic length scale.

The main problem of the Wilsonian approach is that the critical points lay in the strong-coupling regime, so that reliable calculations can be done only by using computers. This has been done in the framework of casual dynamical triangulation (CDT) models \cite{cdt}. In CDT approach $Z_R$ is computed by summing over a special class of folliated triangulations such that all the spacelike $L_\e$ have the same length $l_s$ while the timelike $L_\e$ have a different length $l_t$.

Recently, an alternative approach to the problem of the continuum limit of a discrete quantum gravity (QG) theory was proposed, in the form of the effective action (EA) formalism \cite{mvea,mvea2,scube}. The idea is to apply the EA formalism from quantum field theory (QFT) to discrete path integrals which appear in state-sum models of QG. The initial motivation was to investigate the semiclassical limit of spin foam models, but it turned out that one can also address the issue of the continuum limit.

Spin foam (SF) models, for a review and references see \cite{sfm}, represent a proposal to define $Z_0$ by using a triangulation where the basic variables are the areas $A_\D$ which are determined by finite-dimensional $SU(2)$ representations (spins). The motivation comes from loop quantum gravity (LQG) and $Z_{SF}$ is given as a sum over spins and intertwiners of amplitudes constructed from products of weights for triangles, tetrahedrons and 4-simplices. The weights can be chosen such that $Z_{SF}$ is finite \cite{sffin}, and the corresponding effective action can be shown to be the area-Regge action plus small quantum corrections if the spins, or equivalently the triangle areas, are large \cite{mvea,mvea2}. For geometric configurations, the area-Regge action becomes the Regge action, and as the triangulation is refined, the corresponding Regge action is very well approximated by the Einstein-Hilbert action. Hence the diffeomorphism invariance appears as an approximate symmetry of the SF effective action when the number of simplices in a triangulation is large \cite{mvea}. The same happens in the case of spin-cube models \cite{scube}, which are generalization of SF models such that the edge lengths are independent variables \cite{mv2p}.

In this paper we are going to study the effective action for the theory defined by the path integral $Z_R$ and the measures of the QRC type (\ref{qrcm}). We will assume that the spacetime is fundamentally discrete, so that it will not be necessary to define the $E\to\infty$ limit of $Z_R$. However, we will need the large-$E$ asymptotics of $Z_R$ in order to describe the QG effects for smooth spacetimes. The study of the effective action for $Z_R$ started in \cite{scube}, but there only the power-law measures, which behave as
$$\m (L) \approx \prod_{\e=1}^E (L_\e)^p \,, $$
for  $L$ large and $p\in{\bf R}$, were considered. Also, the effective action was defined in \cite{scube} by using an equation where the positivity of $L_\e$ was not imposed. We will then study the effective action for power-law, exponential and QRC measures and explore the implications of imposing the edge-length positivity in the EA equation.

In section 2 we define a Regge state-sum model and we study the corresponding effective action equation in the one-dimensional case for power-law and exponential measures when the positivity of the edge lengths is imposed. In section 3 we study the same problem in a higher-dimensional case, and derive the conditions for the validity of the semiclassical approximation. In section 4 we show that the exponential measures of the QRC type (\ref{qrcm}) can generate the cosmological constant term in the effective action. In section 5 we discuss the smooth-manifold approximation of the one-loop effective action. In section 6
we study the EA equation when the edge lengths take a discrete set of values and in section 7 we present our conclusions. We also include two appendicies with relevant formulas for the error function and Gaussian sums.

\bigskip
\bigskip
\noindent{\bf{2. Regge state-sum models}}

\bigskip
\noindent We will define a Regge state-sum model by fixing a spacetime triangulation $T(M)$ of a compact 4-manifold $M$. $T(M)$ will be a four-dimensional simplicial complex, and we will label the edges $\e$ of $T(M)$ with non-negative numbers $L_\e$. These numbers will be called edge lengths, since we will require that $L_\e$ for every triangle $\D$ of $T(M)$ satisfy the triangle inequalities. In order to completely specify a Regge state-sum model, we have to specify the set of values of $L_\e$. The simplest choice is the interval $[0,\infty)$, but one can also consider $[a , \infty)$ and $[a , b ]$ intervals, where $a,b >0$. We will also consider the case of a discrete set of values, given by $L_\e = l_0\, n$, $n\in \bf N$ and $l_0 >0$.

Let $L_\e \in [0 , \infty)$. The corresponding state sum can be defined by the Regge path integral (\ref{rpi}), which we write as
\be Z_R = \int_{ D_{E}} \, d^E L \,\mu (L) \exp \left( i{S}_R (L)/l_P^2\right) \,, \label{crss}\ee
where $d^E L = \prod_{\e=1}^E  \, dL_\e $.
The reason that in (\ref{crss}) appears the Planck length is that the GR action $S_{GR}$  is given by $ S_R /G_N$, so that $S_{GR} / \hbar = S_R / l_P^2$.

The measure $\m$ has to be chosen such that $Z_R$ is convergent and that the corresponding effective action allows a semiclassical expansion around the classical limit given by $S_R$. We will first study this problem for the measures
which behave for large $L_\e$ as
 \be \m (L) \approx \prod_{\e=1}^E (L_\e /L_1 )^{p} \, e^{-(L_\e /L_0)^\a} \,,\label{genm}\ee
where $L_0, p$ and $\a$ are parameters to be determined. $L_1$ is an arbitrary length, introduced only to make $\m$ dimensionless. This class of measures is a simplified version of the QRC class (\ref{qrcm}), and (\ref{genm}) will serve as a guide for analyzing the QRC case. We will restrict $\a$ to be non-negative number, so that when $\a > 0$ we will have a finite $Z$. When $\a =0$, we will assume that $p$ is such that $Z$ is finite, see \cite{scube}. 

The effective action $\G (L)$ can be defined by using an integro-differential equation from QFT, see \cite{ea,n}. However, in the QG case the ``field variable'' $L_\e$ does not take all the values from $\bf R$, but $L_\e \ge 0$. In order to see the difference it is sufficient to consider the case of a single variable $L \in [a,b]$. Let us start from the generating functional
$$ Z(J) = \int_{a}^b dL\,\m (L) \exp\left(\frac{i}{\hbar}S(L) + i JL \right) =  e^{\frac{i}{\hbar} W(J)}\,,$$
where $S(L)$ is a $C^{\,\infty}$ function. We define the ``classical field'', also known as the background field, as
$$ \bar L = \frac{1}{\hbar} W'(J)\,,$$ 
where $W(J) = \log Z(J)$.

The corresponding Legandre transformation is given by
$$ \G (\bar L ) = W(J) - \hbar J \bar L \,,$$
so that we obtain 
\be e^{i\G (\bar L)/\hbar} = \int_{a -\bar L}^{b-\bar L} dl\,\m(\bar L+l) \exp\left(\frac{i}{\hbar}[S(\bar L + l) - \G'(\bar L)\,l]\right)\,,\label{gene}\ee
where $l= L - \bar L$ is the ``quantum fluctuation''. In the QFT case $a = -\infty$ and $b = \infty$, which gives the equation
\be e^{i\G ( L)/\hbar} = \int_{-\infty}^{\infty} dl\,\m( L+l) \exp\left(\frac{i}{\hbar}[S( L + l) - \G'( L)\,l]\right)\,.\label{qfte}\ee
In the simplest QG case we have $a = 0$ and $b = \infty$, so that
\be e^{i\G ( L)/\hbar} = \int_{-L}^{\infty} dl\,\m( L+l) \exp\left(\frac{i}{\hbar}[S( L + l) - \G'( L)\,l]\right)\,.\label{qges}\ee

Let us look for a perturbative solution of (\ref{qges}) in the form
\be \G(L) = S(L) + \hbar \G_1 (L) + \hbar^2 \G_2 (L) + \cdots \label{pexp}\ee
when $L \to\infty$ and with a measure which satisfies (\ref{genm}). 
The appearence of a semi-infinite interval of integration in the QG case may change the nature of the perturbative solution. Namely, the QFT perturbative expansion is based on the Gaussian integration formula
\be \int_{-\infty}^\infty e^{-zx^2/\hbar-wx}\, dx = \sqrt{\pi \hbar \over z}\,e^{\hbar w^2 /4z} = \sqrt{\pi\hbar}\, \,e^{-\frac{1}{2}\log z + \hbar w^2 /4z}\,,\label{qfti}\ee
where $Re\, z > 0$ or $Re\,w \ge 0$ if $Re\, z =0$. In the QG case it changes to
\bea \int_{-L}^\infty e^{-zx^2/\hbar -wx}\, dx &=& \sqrt{\pi \hbar \over z}\,e^{\hbar w^2 /4z}\left[\frac{1}{2} + {1\over 2} \,\textrm{erf} \left( L\sqrt{{z\over\hbar}}+ {\sqrt{\hbar}w\over 2\sqrt{z}}\right) \right] \cr
&=& \sqrt{\pi\hbar}\exp {\Big [}-\frac{1}{2}\log z + {\hbar w^2 \over 4z}\cr
&\quad& \quad +{\sqrt{\hbar}e^{-z\bar{L}^2/\hbar} \over 2\sqrt{\pi z}\bar L}\left(1 + O\left({\hbar \over z \bar{L}^2}\right) \right){\Big ]} \,,\label{qgi}\eea
where $\bar L = L + \hbar w/2z$, see the Appendix A. 

The key diference between (\ref{qfti}) and (\ref{qgi}) is the appearence of the non-analytic term in $\hbar$ in (\ref{qgi}), which is given by $\sqrt\hbar\, e^{-z\bar L^2 /\hbar}$. Hence, if this non-analytic term is not supressed for large $L$, we will not have a semiclassical solution for large $L$. Therefore we need 
$$\lim_{L\to\infty}Re\,( z \bar L^2 /\hbar) = +\infty \,,$$ 
where
$$z \bar L^2 /\hbar  = L^2 z + L w  + (w^2 /4z ) \hbar \,.$$

In the QG case we can replace $\hbar$ with $l_P^2$ and 
$$ z/l_P^2 =-iS''(L)/2l_P^2  +  p / L^2 + \a (\a -1)(L_0)^{-\a} L^{\a -2} \,,$$ 
while
$$ w = -i(\G_1' + l_P^2 \G'_2 + \cdots ) \,.$$

Let us assume that $\a$ and $p$ have such values that allow a perturbative solution. In the next section we will show that in the perturbative case
\be \G_1 = -i p \ln (L/L_1) + i (L/L_0)^\a + \frac{i}{2} \log S''(L) \,,\label{olg}\ee
and
$$\G_{n+1}(L) = O(L^{n(\a -2)}) \,.$$
Hence
$$ Re (z\,\bar L^2 /l_P^2) =  \a^2 (L/L_0)^{\a} + {LS'''(L)\over 2S''(L)}+ O( L^{\a -2}) + O(L^{2\a -4}) + O(L^{3\a -6}) + \cdots \,. $$

If $0 < \a < 2$, then
$$Re (z\,\bar L^2 /l_P^2) \approx  \a^2(L/L_0)^{\a} \,,$$
when $L\to \infty$, since $LS'''(L)/S''(L) = O(1)$ due to $S(L) = O(L^2)$ in the GR case,
so that  the non-perturbative terms will be exponentially supressed. In this case we can use the QFT equation (\ref{qfte}) to obtain $\G_n (L)$ for large $L$.

If $\a \ge 2$, we cannot say what is the large $L$ asymptotics of $Re (z\,\bar L^2 /l_P^2)$. However, in the next section we will see that it is possible to answer this question by studying the structure of the perturbative series (\ref{pexp}).

If $\a =0$, then
$$Re (z\,\bar L^2 /l_P^2) \approx  {LS'''(L)\over 2S''(L)} \,, $$
which is a bounded oscilating function. This implies that the non-perturbative terms will not be supressed for large $L$, so that the semiclassical solution does not exist for any value of $p$. This is quite surprising given that the perturbative solution exists for any $p$ in the QFT interval case.

\bigskip
\bigskip
\noindent{\bf{3. Higher-dimensional case}}

\bigskip
\noindent Let us now try to generalize the analysis of the previous section to the case $E > 1$. Let $L=(L_1,\cdots,L_E ) \in D_E$, then we obtain the following integro-differential equation
\be e^{i\G (L)/l_P^2} = \int_{D_E (L)} \, d^E l \, \mu (L + l ) \, \exp \left( i S_R (L+l)/l_P^2 - i\sum_\e \frac{\partial\G}{\partial L_\e }\,l_\e /l_P^2 \right)\,,\label{hde}\ee
where the integration region $D_E (L)$ is a subset of ${\bf R}^E$ obtained by translating $D_E$ by a vector $-L$. 

The main problem with generalizing the $E=1$ results is that we do not know how to calculate exactly the integral
$$I_0 = \int_{D_E (L)} d^E l \exp\left( -\langle l,zl \rangle/l_P^2 + \langle w,l\rangle\right) \,, $$
where $z$ is an $E\times E$ symmetric complex matrix and $\langle x,y\rangle = \sum_{k=1}^E x_k \, y_k $.
A reasonable conjecture is that for large $L$
$$I_0 \approx \int_{C_E (L)} d^E l \exp\left( -\langle l,zl\rangle/l_P^2 + \langle w,l\rangle\right) \,, $$
where
$$C_E(L) = [-L_1 ,\infty)\times \cdots \times [-L_E ,\infty) \,.$$

From this conjecture it follows that
$$I_0 \approx \left({\pi l_P^2 \over 4}\right)^{E/2}(\det z)^{-1/2}\,e^{l_P^2 \langle w, z^{-1} w\rangle/4}\prod_k \left[ 1 + \E\left( {\tilde L_k \sqrt{\l_k}\over l_P} +  {l_P \tilde w_k \over 2\sqrt{\l_k}} \right)\right] \,,$$
where $\l_k$ are the eigenvalues of the matrix $z$ and $\tilde w = U w$, where $U$ is the matrix which puts $z$ in the diagonal form, i.e. $ z = U^{-1} \,diag\, (\l_1 , ..., \l_E ) U$.

Then a semiclassical solution of (\ref{hde}) will exist for the measures satisfying (\ref{genm})
when $L \gg l_P$ and $0 < \a < 2$. As in the $E=1$ case, we cannot determine what happens for $\a \ge 2$, so that we have to use a different method. Let us assume that $\a$ and $p$ are such that the perturbative expansion
\be \G(L) = S_R(L) + l_P^2 \G_1 (L) + l_P^4 \G_2 (L) + \cdots  \label{plex}\ee 
is valid. We will try to derive some restrictions on $\a$ and $p$ from the requirement that 
\be  {l_P^2 |\G_{n+1}(L)| \over |\G_n (L)|} \ll 1 \,\label{scr}\ee 
for $L/l_P \gg 1$ and all $n$. 

The requirement (\ref{scr}) defines the semiclassical expansion, since it implies that the quantum corrections are much smaller than the classical value. A weaker version of (\ref{scr}) is 
\be {l_P^2 |\G_{n+1}(L)|\over |\G_n(L)|} < 1 \label{pr}\ee
for $L /l_P > 1$ and all $n$, and in this case we will consider the solution (\ref{plex}) to be perturbative. Note that the series (\ref{plex}) is an asymptotic series, so that it does not have to converge, and therefore (\ref{pr}) is a way to estimate the region where the perturbative expansion is valid.

Let $\a > 0$ and if the perturbative expansion (\ref{plex}) is valid we can use the approximation $D_E(L) \approx {\bf R}^E$ to solve the equation (\ref{hde}). We obtain
\be \G_1 (L) = i \sum_{\e=1}^E  \left[(L_\e /L_0)^{\a} - p\ln(L_\e /L_1)\right] + \frac{i}{2}\, Tr\,\log S_R''(L) \,,\label{olc}\ee
which is of $O(L^\a)$\footnote{We define $f(x_1,x_2,\cdots,x_n) = O(x^\a)$ if $f(\l x_1,\l x_2,\cdots,\l x_n) \approx \l^\a g(x_1,x_2,\cdots,x_n)$ for $\l\to\infty$.}. The higher-order corrections $\G_n$ can be determined by using the diagramatic technique from QFT, see \cite{k}. These corrections can be evaluated by using the effective action diagrams (EAD), whose $k$-valent vertices ($k \ge 3$) carry the weights $ S_k = i S_R^{(k)}(L)/k!$ and the edges carry the propagator $ G(L) = i( S_R'')^{-1}$. The contributions from a non-trivial path-integral measure can be taken into account if in the formulas for the vertex weights and the propagator we replace $S_R$ by
$$\bar S_R = S_R + i \,l_P^2 \sum_{\e =1}^E \left[(L_\e /L_0)^\a - p\ln(L_\e /L_1)\right]\,.$$

This follows from the EA equation (\ref{hde}), since it can be rewritten as the EA equation with a trivial measure term and the action $\bar S$. The perturbative solution will take the form
$$ \G = \bar S_R + l_P^2 \bar\G_1 + l_P^4 \bar\G_2 + \cdots \,,$$
where $\bar\G_n$ will be given by the EAD with $\bar G$ propagator and $\bar S_k$ verticies. Since
$$\bar\G_n = \G_{n,0} + l_P^2 \bar\G_{n,1} + l_P^4 \bar\G_{n,2} + \cdots \,,$$
we obtain
\be \G = S_R + l_P^2 ( -i\log\m + \G_{1,0} ) + l_P^4 ( \G_{2,0} + \bar\G_{1,1} ) + l_P^6 (\G_{3,0} + \bar\G_{1,2} +\bar\G_{2,1} ) + \cdots \,.\label{pe2}\ee

For example
$$ \G_2 = \langle (S_3)^2 G^3 \rangle + \langle S_4 G^2 \rangle + Res\left[l_P^{-4} \, Tr\log\bar G \right] $$
$$ \G_3 = \langle (S_3)^4 G^6 \rangle + \langle S_3 S_4 G^4 \rangle + \langle S_6 G^3 \rangle + Res\left[l_P^{-6} \left( Tr\log\bar G  + \langle (\bar S_3)^2 \bar G^3 \rangle + \langle \bar S_4 \bar G^2 \rangle\right) \right] $$
and so on. Here $\langle X Y \cdots \rangle$ denotes the sum of all possible contractions of the tensors $X$, $Y$, ..., which is given by the corresponding EAD, see \cite{k}. The residum terms are determined by the formula 
$$Res\,(z^{-n}f(z))= {f^{(n-1)}(0)\over (n-1)!}\,,$$
where $z = l_P^2$.

By using that $S_n = O(L^{2-n})$ and $\bar S_n = O(L^{2-n}) + O( L^{\a-n})$, we obtain
$$\G_{n+1,0} = O(L^{-2n}) \,,\quad \bar\G_{n+1-k,k} = O(L^{k\a - 2n}) \,,$$
where $k=1,2,...,n$. Consequently
$$l_P^{2n}\,\G_{n+1,0} = O((l_P /L)^{2n}) \,,\quad l_P^{2n}\,\bar\G_{n+1-k,k} = O((L/L_0)^{k\a}(l_P /L)^{2n}) \,.$$
Hence
$$ l_P^{2n} \,\G_{n+1} (L) =O\left(\left(l_P / L\right)^{2n}\right) + O\left(\left(l_P / L\right)^{2n} \left(L / L_0\right)^{n\a} \right)  $$
$$= O\left(\left(l_P / L\right)^{2n}\right) 
+  O\left(\left(L / L_{s} \right)^{n(\a -2)}\right)\,,$$
where
\be L_{s} = \left( L_0^{\a}/l_P^{2}\right)^{1\over \a -2}\,,\label{nls} \ee
is a new length scale which together with $l_P$ will determine the validity of the semiclassical expansion. 

By using the criterion (\ref{scr}) we obtain that the perturbative expansion (\ref{pe2}) will be semiclassical if
\be l_P^2 /L^2 \ll 1 \,,\quad (L/L_s )^{\a -2} \ll 1 \,.\label{scc}\ee 
The condition (\ref{scc}) will be satisfied if
$L \gg l_P$ and $L \gg L_{s}$ for $\a < 2$, while for $\a > 2$ we need that
$ l_P \ll L \ll L_{s} $.

When $\a = 2$, we have
$$ l_P^{2n} \,\G_{n+1} (L) =O\left(\left(l_P / L\right)^{2n}\right) + O\left(\left(l_P / L_0\right)^{2n} \right)\,, $$
so that the series (\ref{pe2}) will be semiclassical for $L_0 \gg l_P$ and $L \gg l_P$. 

Note that the $\G_1$ takes imaginary number values, and the higher-order quantum corrections $\G_n$ will in general take complex number values. The same happens in QFT, and since we want to have a real effective action, we have to restrict our complex solution to real values. In QFT this is done by using the Wick rotation. However, the Wick rotation requires a flat background metric, and introducing a background metric in QG at the fundamental level is contrary to the purpose of QG. Fortunatelly, it was observed in \cite{mvea} that in QFT the Wick rotation is equvalent to the transformation
\be \G \to  Re\,  \G (L) \pm  Im\,  \G (L)\,.\label{wrqg}\ee
Since the transformation (\ref{wrqg}) is metric-independent, it can be used in QG to define a real effective action. The sign ambiguity can be fixed by an experimental input, see the next section.

\bigskip
\bigskip
\noindent{\bf{4. Cosmological constant measures}}

\bigskip
\noindent When $\a > 2$ the expansion (\ref{pe2}) will be semiclassical if $l_P \ll L \ll L_s$, so that we need that $L_s \gg l_P$, which is satisfied if $L_0 \gg l_P$. The interesting case is $\a = 4$, because the quantum Regge calculus measures (\ref{qrcm}) are of this type. More generally, one can consider the measures which satisfy
\be \log \m = O(L^4) \,,\label{gccm}\ee 
and $\log\m < 0$ for $L$ large .

Let us consider the following PI mesure
\be  \m_c(L) = \exp\left( - \sum_{\s=1}^N V_\s (L) /L_0^4 \right) = \exp\left( - V_4 (L) /L_0^4 \right) \,.\label{ccm}\ee
This measure is a special case of the measure (\ref{qrcm}), and we have taken $p=0$ and $q=0$ for the sake of simplicity.
Given that $V_\s (L) = O(L^4)$ for large $L$, we have 
$$ \log \m_c (L) = O((L/L_0)^4) \,,$$ 
so that we can apply the same reasoning when calculating the perturbative effective action as in the case of the $\a = 4$ measure (\ref{genm}) with $p=0$. 

Namely, if we substitute $D_E (L)$ in (\ref{hde}) by ${\bf R}^E$ we obtain
\be \G_1 (L) = i {V_4 (L)\over L_0^4} + \frac{i}{2}\, Tr\,\log S_R''(L) \,,\label{cmf}\ee
so that $\G_1 (L) = O((L /L_0)^4)$ and
$$\bar S_R = S_R + i \,l_P^2 V(L)/L_0^4 \,. $$
Hence the formulas from the previous section give
$$ \G = \bar S_R + l_P^2 \bar\G_1 + l_P^4 \bar\G_2 + \cdots  = S_R + l_P^2 \G_1 + l_P^4 \G_2 + \cdots \,,$$
where 
$$ l_P^{2n} \,\G_{n+1} (L) = O\left(\left(l_P / L\right)^{2n}\right) 
+  O\left(\left(L / L_{s} \right)^{2n}\right)\,,$$
and
\be L_{s} =  {L_0^{2}\over l_P}\,.\label{ccl} \ee
Therefore the effective action will have a semiclassical expansion if
\be l_P \ll L_\e \ll {L_0^{2}\over l_P} \,,\label{ccsr}\ee
which is satisfied for $L_0 \gg l_P$.

If we define the physical effective action as 
$$S_{eff} = {1\over G_N}\,\left(Re\,\G \pm Im\,\G\right) \,,$$
where the $\pm$ sign corresponds to a negative/positive cosmological constant, we then obtain from (\ref{cmf})
\be S_{eff} = {S_R \over G_N} \pm {l_P^2 \over G_N \, L_0^4}\,V_4  \pm {l_P^2\over 2G_N}\,Tr \log S''_R  + O(l_P^4) \,.\label{eacc}\ee
Hence the second term in (\ref{eacc}) can be interpreted as the cosmological constant term with the value of the cosmological constant given by
\be \L = \mp {l_P^2 \over 2 \, L_0^4} = \mp {1\over 2 L_s^2} \,. \label{vcc}\ee

Note that the value (\ref{vcc}) will be very small in units of $l_P^{-2}$, since 
$$l_P^2 |\L| = \frac{1}{2}\left({l_P \over  L_0 }\right)^4$$ 
and (\ref{ccsr}) gives $l_P / L_0 \ll 1$. Therefore we have a mechanism to generate a small cosmological constant from the PI measure (\ref{ccm}), as a first-order quantum correction. If we define $L_\L = 1/\sqrt{\L}$, then the observed value for $\L$ gives $ L_\L \approx 10^{26}\,m$. Since $L_s = L_\L /\sqrt{2}$, we get $L_0 \approx 1\,mm$ so that $ l_P /L_0 \approx 10^{-32}$.

Observe that $\G_3 (L) = O(L^4)$, so that one can have an $O(l_P^6 /L_0^8)$ correction to the CC value (\ref{vcc}). Hence the exact value of the cosmological constant will be given by
\be  \L = \mp {l_P^2 \over 2 \, L_0^4}\left( 1 + c_3 \,{l_P^4 \over L_0^4 }\right) \,,\label{evcc}\ee
where $c_3$ is a numerical constant of $O(1)$. Since $ l_P /L_0 \approx 10^{-32}$ for the observed value of the cosmological constant, this correction can be neglected.

In the case of a QRC measure with $p\ne 0$ or $q\ne 0$ we will obtain the same formulas for the cosmological constant. This is because the corresponding terms in the $\log\m$ term are of subleading order in $L$ with respect to the $V_4 (L)$ term.

\bigskip
\bigskip
\noindent{\bf{5. Smooth-manifold approximation}}

\bigskip
\noindent When the number of the edges $E$ is large, one can obtain a smooth-manifold approximation of $\G_n (L)$ terms. Here we will only discuss the the leading quantum corrections, which are given by $\G_1(L)$. We will then look for a smooth-manifold approximation of
$$ \G (L) \approx S_R (L) \pm l_P^2 \G_1 (L) \,, $$
where
\be \G_1 (L) = -\log \m (L) + {1\over 2} \,Tr \log S_R''(L) \,. \ee

For the Regge action it is known that
\be S_R (L) \approx \int_M d^4 x \,\sqrt{|g|}\,R(g)  \,,\label{rah}\ee
when $E\to\infty$ and $|g|=|\det \,g|$. Therefore the Einsten-Hilbert action is a good approximation for the Regge action when $E\gg 1$.

In the case of an exponential measure (\ref{genm}), the terms in $\log\m$ give
$$ \sum_{\e=1}^E \,( L_\e /L_0  )^\a \approx \, \int_M d^4 x \,\sqrt{|g|}\, {\cal F}_\a [g(x)] \,,$$
and
$$ \sum_{\e=1}^E \,\log (L_\e /L_1) \approx \, \int_M d^4 x \,\sqrt{| g|}\, {\cal G} [g(x)] \,,$$
for $E$ large. However, we do not know whether diffeomorphism-invariant functionals ${\cal F}_\a$ and ${\cal G}$ exist and what is their form.

The same problem appears for a QRC measure (\ref{qrcm}) with $p \ne 0$. Also when $q\ne 0$, the $\log\m$ term contains 
$$ \sum_{\s=1}^N \,\log V_\s (L) \approx \,  \int_M d^4 x \,\sqrt{| g|}\, {\cal H} [g(x)] \,,$$
and it is not known wheather a diffeomorphism-invariant functional $\cal H$ exists. Only in the case when $p=q=0$, which corresponds to a CC measure, we know what is the smooth-manifold approximation of the $\log\m$ term. It is proportional to the 4-volume of $M$, since
\be V_4 (L) \approx \,  \int_M d^4 x \,\sqrt{| g|} \,.\label{fv}\ee

Note that there is a conjecture that $p=1$ QRC measures are diffeomorphism invariant measures, see \cite{h}. If true, this conjecture implies that diffeomorphism-invariant functionals $\cal G$ and $\cal H$ exist. Also note that a CC measure (\ref{ccm}), which has $p=0$, can be considered as a diffeomorphism invariant measure.

As far as the trace-log term is concerned, its continuum approximation can be obtained by using the effective field theory approach, see \cite{don, burg, wein}. If we  introduce a cutoff scale $L_c$ such that 
\be L_\e \ge L_c \gg l_P \,,\label{ctd}\ee
then for $E \gg 1$ we will have
\be Tr (\log S_R''(L) ) \approx \, \int_M d^4 x \sqrt{|g|} \left[ a(L_c)  R^2 + b(L_c) R_{\m\n} R^{\m\n} \right] \,,\label{trl}\ee
where the dimensionless functions $a(L_c)$ and $b(L_c)$ can be determined by using the one-loop EA diagrams from QFT which have the momentum ultra-violet cutoff $\hbar /L_c$. Note that $L_c$ can be chosen to be the minimal distance for which we know that perturbative QFT is applicable. From the LHC experiments we know that $L_c \le 10^{-20}\,m$.

In the case of a CC measure, by combining (\ref{rah}),(\ref{fv}) and (\ref{trl}) we obtain 
\be \G(L) \approx \int_M d^4 x \sqrt{|g|} \left[ R - \L + a(L_c)\, l_P^2 \,R^2 + b(L_c)\,l_P^2 \, R_{\m\n} R^{\m\n}\right]\,,\ee
where $L$ satisfies (\ref{ctd}), $E\gg 1$ and $\L$ is given by (\ref{vcc}).

\bigskip
\bigskip
\noindent{\bf{6. Discrete-length Regge models}}

\bigskip
\noindent Let us now analyze the effective action for discrete-length Regge state sum models. In this case we can write $L_\e = \g n_\e l_P$ where $n\in {\bf N}$ and $\g > 0$. Then
$$ Z_R = \sum_{n \in N_{\g,E}}\, \m(L(n)) \, \exp \left( iS_R (L(n))/l_P^2\right) \,, $$
where $N_{\g,E}$ is a subset of ${\bf N}^E$ such that $\g n l_P \in D_E$. 

The effective action equation is given by
\be e^{i\G (L)/l_P^2} = \sum_{ n \in N_{\g,E}} \, \mu (L + l(n)) \, \exp \left( iS_R (L+l(n))/l_P^2- i\sum_\e \frac{\partial\G}{\partial L_\e }\,l_\e(n) /l_P^2\right) \,,\label{dlr}\ee
where $l_\e(n) = \g l_P n_\e - L_\e$. Note that the variable $L$ in (\ref{dlr}) is actually a quantum expectation value of $L$, which we have denoted as $\bar L$, see (\ref{gene}). We will refere to this $L$ as the background $L$, and it can take any value in ${\bf R}^E$. However, for the sake of simplicity we  will consider the backgrounds such that $ L / \g l_P \in {\bf Z}^E$, so that $l / \g l_P \in {\bf Z}^E$. Otherwise $l_\e /\g l_P = m_\e + x_\e$ where $m_\e \in {\bf Z}$ and $x_\e \in (0,1)$.

One expects to obtain the same results for the semiclassical solution of (\ref{dlr}) as in the continious case. However, there is an obstruction, due to the fact that
$$ \sum_{m=-k}^\infty f(m) \ne \int_{-k}^{\infty}  f(l) \,dl \,.$$
In our case this problem appears when computing the one-loop correction, which is given by the logarithm of
$$\sum_{m\in \, {\bf Z}^E} \exp\left(\frac{i}{2}\langle \g m \,,S_R''(L)\g m\rangle \right) \,.$$
Since
$$ \sum_{m\in \,{\bf Z}} \exp(iam^2 ) \ne \sqrt{\frac{i\pi}{ a}}\,,$$
we cannot use the Gaussian integral approximation. However, one can show that
$$ \sum_{m\in \,{\bf Z}} \exp(iam^2 ) \approx \sqrt{\frac{i\pi}{ a}}\,,$$
for $a \to 0$ (see the Appendix B). Hence
\be \sum_{m\in\, {\bf Z}^E} \exp\left(\frac{i}{2}\langle m,\g^2 S_R''(L)\, m\rangle \right)
\approx \left(2i\pi \right)^{E/2}(\det( \g^2 S_R''(L))^{-1/2} \,,\label{gs}\ee
only if the entries of the Hessian matrix $S''_R (L)$ satisfy 
\be \g^2 |S_R''(L)| \ll 1 \,.\label{dsca}\ee 
Since $S_R''(L) = O(1)$, we need $\g^2 \ll 1$ which implies $\g \ll 1$. 

Therefore the semiclassical approximation will be valid only if the spectrum gap is much smaller than $l_P$. 
This is a surprising result, since it implies that in the natural case when the spectrum gap is of order $l_P$, which corresponds to  $\g\approx 1$, one cannot solve the EA equation (\ref{dlr}) perturbatively. Even if we abandon
the positivity of $L+l$, and replace $D_E (L)$ with ${\bf R}^E$, the result (\ref{dsca}) holds.

The requirement (\ref{dsca}) can be also applied to the semiclassical approximation of the effective action for spin foam models. In the spin foam case, instead of the edge lengths, we have the triangle area variables $j\,l_P^2 $, such that $j \in ({\bf N}/2)^F$ and $S_R (L)/l_P^2 \to S(j)$ where
$$ S (j) \approx \sum_{f=1}^F j_f \,\theta_f (j) \,, $$
for $j_f \gg 1$ and $\theta(j) = O(1)$, see \cite{mvea}.  Hence $\g = 1/2$ in the spin foam case. However, there is no problem for the semiclassical approximation, since the Hessian satisfies $S''(j) = O(1/j)$. Therefore $|S''(j)| \ll 1$ so that
$$ \sum_{m\in\, {\bf Z}^F} \exp\left(\frac{i}{8}\langle m, S''(j)\, m\rangle \right)
\approx \left(8i\pi \right)^{F/2}(\det( S''(j))^{-1/2} \,.$$

Note that the spin foam analog of the $D_E (L)$ integration region is 
$$D_F (j) = [-j_1 ,\infty) \times \cdots \times [-j_F ,\infty) \,.$$
In order to have a perturbative expansion of the SF effective action for large $j_f$ we need to modify the standard SF measure
$$ \m (j) = \prod_{f=1}^F \dim j_f = \prod_{f=1}^F (2j_f +1) \,, $$
by including an exponentially damping term. For example, the measure
$$ \tilde\m (j) = \prod_{f=1}^F (2j_f + 1)\, e^{-(j_f)^\a} \,,$$
where $\a > 0$, will allow a semiclassical solution. In order to have an explicit LQG interpretation of the modified measure, one can replace $(j_f)^\a$ with $(j_f (j_f +1))^{\a/2}$, where $j_f (j_f +1)$ is an $SU(2)$ Casimir operator eigenvalue.

\bigskip
\bigskip
\noindent{\bf{7. Conclusions}}

\bigskip
\noindent Our analysis implies that the power-law measures which satisfy (\ref{genm}) for $\a =0$, do not allow a semiclassical solution of the EA equation when the integration region is consistent with positivity of the edge lengths. However, the exponential
measures which satisfy (\ref{genm}), or more generally (\ref{gccm}), allow the semiclassical solution provided that $L$ satisfies the conditions (\ref{scc}). In this case the semiclassical solution can be obtained by replacing the integration region $D_E (L)$ with ${\bf R}^E$. 

The case $\a = 4$ includes the quantum Regge calculus measure (\ref{qrcm}) and the CC measure (\ref{ccm}). The condition for validity of the semi-classical expansion (\ref{scc}) implies that the edge lengths cannot be too small nor too large. The upper bound is given by $L_0^2 /l_P $,  and since this length must be much larger than the Planck length, one obtains that $L_0 \gg l_P$. This is the reason why the the corresponding cosmological constant, given by (\ref{evcc}), is very small in $l_P^{-2}$ units. 

Note that the effective action used for spin-foam and spin-cube models in \cite{mvea,mvea2,scube} was defined by using the EA equation where the integration region was chosen to be ${\bf R}_+^E$. This integration region gives the same one-loop result as the $D_E(L)$ region for large $L$. Our results imply that at higher loops a better approximation would be to use the QFT integration region ${\bf R}^E$, since $D_E (L) \approx {\bf R}^E$ for large $L$. Also, an exponantial measure should be used in order to have a perturbative solution when the positivity of areas and lengths is imposed.  

An exciting developement is that the measures (\ref{qrcm}) and (\ref{ccm}) can generate the cosmological constant term. We have obtained an exact formula (\ref{evcc}) for the value of cosmological constant, which is practically the same as the first-order approximation (\ref{vcc}), since the semiclassical approximation requires that $L_0 \gg l_P$. This also insures that the corresponding cosmological constant will take a very small value in the units of $l_P^{-2}$. Hence we have a mechanism to generate a very small cosmological constant as a quantum gravity effect. It remains to be seen what is the contribution of the matter sector to the cosmological constant value. If the matter contribution is for some reason small or zero, one would have a theory with a naturally small cosmological constant. The matter contribution to the cosmological constant will also resolve the sign ambiguity in (\ref{wrqg}), since we know that the observed cosmological constant value is positive.

Another surprising result of our approach is that the validity of the semiclassical approximation in the case of discrete $L$ requires that the spectrum gap is much smaller than $l_P$, since the entries of the Hessian matrix satisfy $S_R''(L) = O(1)$. Hence the natural case where $L_\e$ is an integer multiple of $l_P$ requires a nonperturbative solution of the EA equation. In the case of spin foam models, the triangle area is similarly an integer multiple of $l_P^2$ for large spins, but there is no problem with the semiclassical approximation since the Hessian satisfies $S''(j) = O(1/j)$ for large spins $j$, so that the Gaussian sums can be approximated with Gaussian integrals.

It will be interesting to see how our analysis will change for GR with a classical cosmological constant term. The perturbative analysis will be similar to the one performed here, with a difference being in the shift in the order of $l_P^2$ where the $O(L^4)$ perturbation terms contribute. 

The effective action equation can be solved perturbatively for $L_\e \gg l_P$, which is the semiclassical regime. An important problem is how to solve the EA equation for $L_\e \approx l_P$, which is the deep quantum regime. In this region the perturbation theory fails, and one has to find an alternative method. A promissing approach is to use the fact that the effective action is also the generating functional for the one-particle-irreducible (1PI) Green's functions, so that
\be \G (L) = \sum_{\e,\e'} \tilde\G_2 (\e ,\e') L_\e L_{\e'} + \sum_{\e,\e',\e''} \tilde\G_3 (\e, \e' ,\e'') L_\e L_{\e'} L_{\e''} + \cdots \,, \label{npe}\ee
where $\tilde\G_n (\e)$ is the 1PI part of the $n$-point Green's function
$$ G (\e_1,...,\e_n) = {1\over Z_R}\int_{D_E} d^E L \,\m (L) \, L_{\e_1} \cdots L_{\e_n}\, e^{iS_R(L)/l_P^2} \,.$$
This integral can be calculated numerically, which can be used to obtain the expansion (\ref{npe}). One can also study the non-perturbative effects by using the mini-superspace approximation $L_1 = \cdots = L_E = L$ and the corresponding EA equation. 

Also note that for the exponential mesures with $\a > 2$ there will be
a maximal length $L_s$ for which the semiclassical approximation is valid, so that in this case there may be non-perturbative quantum effects at large distances $L \approx L_s$.

The fact that there is a minimal ($l_P$) and a maximal ($L_s$) length for which the semiclassical approximation is valid, raises the question of the relation of $l_P$ and $L_s$ to the minimal and the maximal length in the spectrum of $L_\e$. Formally, one can choose any interval $[a,b]$ for $L_\e$ of the state-sum model, where $0\le a < b$. In the case $a > l_P$ and $b < L_s$ one will have a QG theory with purely perturbative QG effects. However, a more interesting case is $a \le l_P$ and $b \ge L_s$, since in such a theory one can have non-perturbative QG effects for small and large distances.

Note that the knowledge of the effective action is not sufficient for a complete QG theory. The concept of an effective action only makes sense for spacetimes whose topology is $\S \times [0,1]$, where $\S$ is a 3-manifold.
We also need a wavefunction which can be associated to a cup-manifold $C(\S)$, where $C(\S)$ is a compact 4-manifold whose boundary is $\S$. This is essentially the Hartle-Hawking wavefunction \cite{hh}, and it would be interesting to develop a state-sum quantum cosmology theory based on these concepts.

\bigskip 
\bigskip
\noindent{\bf Acknowledgments}

\bigskip
\noindent I would like to thank M. Vojinovi\'c and M. Oliveira for discussions. This work has been partially supported by the FCT projects PEst-OE/MAT/UI0208/20 11 and EXCL/MAT-GEO/0222/2012.

\newpage
\noindent{\bf Appendix A}

\bigskip
\noindent Consider the following integro-differential equation
$$ e^{i\G(L)/\ve} = \int_{-L}^\infty dl \, e^{i[S(L+l) - \G'(L)l]/\ve} \,,$$
where $S(L)$ is a $C^{\,\infty}$ function, $L > 0$ and $\ve$ is a small parameter. We want to solve it perturbatively in $\ve$ as
$$ \G(L) = S(L) + \sum_{n>0} \ve^n \G_n (L) \,,$$
up to an additive constant.

Since
$$ S(L+l) = S(L) + \sum_{n>0} S_n(L) \, l^n \,,$$
were $S_n (L) = S^{(n)}(L)/n!$, we  obtain
$$ \G_1 + \ve \G_2 + \ve^2 \G_3 + \cdots  = (-i)\log \int_{-L}^\infty dl \exp\left[{i\over\ve}S_2 l^2 - i\bar\G_1' l + {i\over\ve}\sum_{n>2}S_n l^n  \right] \,, \quad (A.1)$$
where $ \bar\G_1 = \G_1 + \ve\G_{2} + \ve^2 \G_{3}+\cdots$.

The integral in (A.1) is of the type
$$ I = \int_{-L}^\infty dl\, e^{-z l^2 +w l} \exp \left(\sum_{n>2}s_n \, l^n \right)\,,$$
which we rewrite as
$$ I = \int_{-L}^\infty dl \, e^{-z l^2 + wl }\left(1 +  \sum_{n>2}\hat s_n \, l^n\right) \,.$$
Hence we will need the integrals
$$ I_n = \int_{-L}^\infty dl \, e^{-z l^2 +wl }\, l^n \,,$$
which can be calculated by differentiating $I_0$ wrt $w$. It is easy to show that
$$ I_0 = \sqrt{\pi\over 4z}\,e^{w^2 \over 4z} \left[ 1+ \E\left(L\sqrt{z} + {w\over 2\sqrt{z}}\right) \right] \,,$$
where
$$\E(x) = {2\over\sqrt{\pi}}\int_0^x e^{-t^2}\,dt \,.$$
The domain of the error function can be extended to any complex number $z$ by using the Taylor expansion
$$ \E(x) = {2\over\sqrt{\pi}}\sum_{n=0}^\infty {(-1)^n x^{2n+1}\over (2n+1)n!} \,.$$ 

For large $x$ we can use
$$ \E(x) = 1 + {e^{-x^2}\over x\sqrt{\pi}}\left( 1+ \sum_{n=1}^{N-1} {(-1)^n (2n-1)!!  \over 2^n x^{2n}} + R_N (x)\right)\,,\quad (A.2)$$ 
where $R_N(x) = O(x^{-2N})$. If $x$ takes complex values, we can use (A.2) for large $|x|$ and $|arg(x)| < 3\pi/4$ \cite{abs}. 
For $|arg(x)| < \pi/2$
$$ R_N (x) = {(-1)^N (2N-1)!!  \over 2^N x^{2N}} \,\theta \,,$$
where 
$$ \theta = \int_0^\infty e^{-t}(1+t/x^2)^{-N -1/2} \,dt  \,.$$
For $|arg(x)| < \pi/4 $ one has $|\theta| < 1$, see \cite{abs}.

\bigskip
\bigskip
\noindent{\bf Appendix B}

\bigskip
\noindent Let
$$ S(a) = \sum_{n=0}^\infty e^{-an^2} \,, $$
where $a>0$. It was shown in \cite{thas} that as $a\to 0$
$$ S(a) = \sqrt{{\pi\over 4a}} + \frac{1}{2}\,e^{-a/4} \left[ {\sinh\sqrt{a}\over\sqrt{a}}- \sum_{n=0}^N c_n \, a^{n +1/2}\,H_{2n+1}(\sqrt{a}/2)\right] + O(a^{N+3/2}) \,,$$
where 
$$ c_n = {(2^{2n+1}-1)B_{2n+2}\over 2^{2n}(2n+2)!} \,,$$
$B_n$ are Bernoulli numbers and $H_n (x)$ are Hermite polynomials. 

If $a<1$, then in the limit $N\to\infty$ we obtain
$$ S(a) = \sqrt{{\pi\over 4a}} + \frac{1}{2}\,e^{-a/4} \left[ {\sinh\sqrt{a}\over\sqrt{a}}- \sum_{n=0}^\infty c_n \, a^{n +1/2}\,H_{2n+1}(\sqrt{a}/2)\right] \,.$$

Let $R(a)= S(a) - \sqrt{{\pi\over 4a}}$, then
$$R(a) = \sum_{n=0}^\infty r_n \, a^{n/2} \,,$$
for $a < 1$. We can now define a complex function 
$$R(z) = \sum_{n=0}^\infty r_n \, z^{n/2} \,,$$
for $|z|<1$. Consequently we can define
$$ S(z) = \sqrt{\frac{\pi}{4 z} } + R (z) \,,$$
for $0<|z|<1$, so that
$$S (-ia) \approx \sqrt{\frac{i\pi}{4a} }$$ 
as $a\to 0$.

\end{document}